\documentclass{aa}
\usepackage{graphics}
\usepackage{txfonts}

\newcommand{\nodata}{---}

\newcommand{\ngc}{NGC\,2359}
\newcommand{\hd}{HD\,56925}
\newcommand{\low}{$1 \rightarrow 0$}
\newcommand{\hig}{$2 \rightarrow 1$}
\newcommand{\kms}{km s$^{-1}$}

\newcommand{\hi}{{\sc Hi}}

\newcommand{\htwo}{H$_2$}
\newcommand{\lmech}{L$_{\rm mech}$}
\newcommand{\msun}{M$_\odot$}
\newcommand{\mlr}{M$_\odot$ yr$^{-1}$}
\newcommand{\tk}{T$_{\rm K}$}

\newcommand{\vlsr}{V$_{\rm LSR}$}
\newcommand{\co}{$^{13}$CO}
\newcommand{\risot}{R$_{\rm isot}$}
\newcommand{\rexc}{R$_{\rm exc}$}
\newcommand{\otf}{{\it on-the-fly}}
\newcommand{\lvc}{{\it lvc}}
\newcommand{\ivc}{{\it ivc}}
\newcommand{\hvc}{{\it hvc}}

\begin{document}


\title{Shocked gas layers surrounding the WR nebula NGC\,2359}

\author{J.~R.~Rizzo\inst{1,2}
\and
J.~Mart\'{\i}n-Pintado\inst{3}
\and
J.-F.~Desmurs\inst{2}}

\institute{Departamento de F\'{\i}sica, Universidad Europea de Madrid, 
Urb.\ El Bosque, Tajo s/n, E-28670 Villaviciosa de Od\'on, Spain
\and
Observatorio Astron\'omico Nacional, Aptdo. Correos 1143, E-28800
Alcal\'a de Henares, Spain
\and
Departamento de Astrof\'{\i}sica Molecular e Infrarroja, Instituto de 
Estructura de la Materia, CSIC, Serrano 121, E-28006 Madrid, Spain
}

\offprints{J.\ R.\ Rizzo,\\
\email{jricardo.rizzo@fis.cie.uem.es}}

\date{Received ; accepted }

\titlerunning{Molecular shocks in NGC\,2359}
\authorrunning{Rizzo et al.}


\abstract{
\ngc\ is a Wolf-Rayet (W-R) nebula partially bound by a rather dense and warm 
molecular cloud. We present the results derived from CO and $^{13}$CO fully 
sampled maps of the molecular material with angular resolutions up to 
12$\arcsec$. We have detected three different velocity components, and 
determined their spatial distribution and physical properties. The kinematics, 
morphology, mass and density are clearly stratified with respect to the W-R star. 
These features allow us to learn about the recent evolutionary history of \hd, 
because the multiple layers could be associated to several energetic events 
which have acted upon the surrounding circumstellar medium. Hence, a careful 
study of the different shockfronts contain clues in determining the present 
and past interaction of this evolved massive star with its surroundings.
From the analysis of the mass-loss history in massive stars like \hd, we 
suggest that the multiple layers of shocked molecular gas are likely to
be produced during the earlier LBV phase and/or the actual W-R stage of \hd.
}

\maketitle


\section{Introduction}

\ngc\ is one of the archetypal wind-driven Wolf-Rayet (W-R) nebula
(Chu et al.\ 1983). It is an optical nebula excited by the WN4 star
\hd\ ((RA, Dec.)$_{2000}$ = 
($07^{\rm h}18^{\rm m}29\fs13,-13\degr13\arcmin01\farcs5$)), which
is catalogued as WR7 by van der Hucht (2001). \ngc\ 
has an almost spherical morphology, and contains some
N-enriched filaments inside (Esteban et al.\ 1990). The
southern part of the nebula --which contains most of the associated 
molecular gas-- has been observed in the recent years. 

St-Louis et al.\ (1998) have detected the \htwo\ 1--0 S(1)
emission in coincidence with H$\alpha$, but they could not
disentangle the dominant excitation mechanism of the molecular
hydrogen. Carbon monoxide was firstly detected by Schneps et al.\
(1981), and subsequently mapped in the J = \low\ and
\hig\ lines by Rizzo et al.\ (2001a, Paper I) and by Cappa et al.\
(2001). While Cappa et al.\ focussed in
determining the total neutral mass and in the energetics, Paper I
has concentrated on the interaction processes between the nebula
and the surrounding molecular gas.

In Paper I, it was shown that the ambient molecular cloud,
observed at a velocity in the Local Standard of Rest (\vlsr) of 67 
\kms, was accelerated to \vlsr\ = 53--54
\kms\ by the stellar wind from \hd.
It was also established the presence of a rather
hot region just coincident with the \htwo\ vibrationally excited
emission surrounding the southern
optical nebula, as well as a high-opacity zone adjacent to the 
cold quiescent cloud. Observations of \co\ have revealed another 
striking feature: in the region of
higher intensity, the \co\ J = \low\ line peaks at 54--55 \kms, 1--2 
\kms\ higher than the CO J = \low\ and the J = \hig\ lines.
It is worth mentioning that, at the same position, the metastable
(1,1) and (2,2) lines of ammonia (NH$_3$) were
detected for the first time in a W-R nebula (Rizzo et al.\ 2001b). From
the NH$_3$ data, a rotational temperature of 30\,K and a relatively
high abundance of $10^{-8}$ were determined.

These observational findings indicate the nature of
the interaction with the ambient material during different evolutive 
stages from the hot star. The molecular cloud has been exposed to the UV 
radiation from the hot star and to the shocks produced by the stellar 
winds from the W-R star and the mass ejections from previous stages. 
According to the 
hydrodynamical models of Garc\'{\i}a-Segura et al.\ (1996a, 1996b),
several shocks are expected at a few pc from a massive evolved star, 
as a consequence of the time-dependent stellar winds of the post-main 
sequence evolution, mainly the red supergiant (RSG) and the luminous blue 
variable (LBV) phases.
Hence, the molecular cloud surrounding \ngc\ provides us with
an excellent laboratory to study the interaction between the massive stars
and their environment. W-R stars will disturb the circumstellar 
medium (CSM), generating an ionized nebula surrounded by a shocked region. 
In particular, the NH$_3$ would be a good molecule to disentangle the effect
of the radiative field and the shock in the environment of W-R stars. 
The relatively high abundance of NH$_3$ strongly suggests a shock chemistry in 
the outer part of the cloud, since it would be released from the dust 
grain mantles by non-dissociative shocks (Flower \& Pineau des For\^ets 
1994; Rizzo et al.\ 2001b).

A detailed knowledge of the spatial and kinematical distributions
of the CO isotopes would allow us to know with more precision the physical
conditions and the kinematical features in this scenario,
together with an approach to the knowledge of the dominant mechanisms of
excitation (UV fields or shocks). In order to satisfy these goals, we 
combine in this paper high angular resolution CO J = \low, CO 
J = \hig\ and \co\ J = \low\ maps
with high sensitivity spectra of CO, \co\ and C$^{18}$O taken at selected
positions. This observational study reveals the presence of hot
dense regions, warm regions and, more conclusively, the stratified distribution 
of different kinematical components, associated to several shockfronts
which have been acting in the recent past of \hd.


\section{Observations}

We have used the IRAM 30-m radiotelescope at Pico Veleta (Spain), during
two observing runs in June and August, 2001. We have performed two 
different observing modes, namely \otf\ maps and single position 
observations.

The \otf\ maps were obtained for the CO and \co\ J\,=\,\low, and the CO 
J\,=\,\hig\ lines simultaneously. We have mapped a $240\arcsec \times 
120\arcsec$ area toward the southern part of \ngc. The scanning directions were 
orthogonal in the equatorial system, e.g., at constant right ascension and 
constant declination. Row spacing between scans were $6\arcsec$, and 
therefore all spectral lines maps are fully sampled. Individual 
maps were later combined using the GILDAS package. 
One receiver was tuned for each J = \low\ line, while the CO 
J\,=\,\hig\ line was observed simultaneously by two different receivers. 
A 256-channels filter bank with 100 kHz of channel resolution and a 2048-channel 
autocorrelator were used 
as backends, providing velocity resolutions between 0.1 and 0.3 \kms. 
Typical {\it rms} was 0.3\,K for all the observed lines. All the 
temperatures throughout this paper are referred to an atmosphere corrected 
scale, e.g., T$_{\rm A}^*$.

The single position observations were performed along a strip of 
particular interest in position-switching mode. We observed the 
J = \low\ and J = \hig\ lines of CO 
and \co\ simultaneously, with integration time between 15 and 40 minutes. 
As a byproduct, the C$^{18}$O J\,=\,\low\ line was also observed with the 
same receiver 
than the \co\ J\,=\,\low\ line. We used the same backend as for the 
\otf\ maps, and hence we had identical velocity resolutions. The achieved 
{\it rms} was better than 50 mK in all cases.

Calibration was performed during the observations and later checked up on 
standard sources (Mauersberger et al. 1989). The HPBW was $22\arcsec$ for the CO 
J = \low\ line, $11\arcsec$ for the CO J = \hig\ line, $24\arcsec$ for the 
\co\ J = \low\ line and $12\arcsec$ for the \co\ J = \hig\ line.


\section{Results}

\subsection{On-the-fly maps}

The southern part of the nebula was mapped in the CO J =
\low, J = \hig\ and \co\ J = \low\ lines simultaneously. In Paper I,
it has been shown that this region contains the most intense and
interesting CO features, between 52 and 67 \kms. The  
maps in Fig.~1 show the integrated emission of these lines
from 50 to 56 \kms. Figure 1a shows the whole CO J = \hig\ emission from
Paper I, superimpossed to an optical image of the nebula. The same
figure also indicates the region mapped in this work.
Figures 1b-1d show the results of these data in the three spectral lines
already mentioned. We can see that the molecular cloud perfectly matchs 
the southern border of the optical nebula. The CO and \co\ emission
are confined to layers which are less than $70\arcsec$ wide.

The overall emission maps depicted in Fig.~1 do not show
significant differences among the observed spectral lines.
However, a number of new features appear when the emission is analysed in
smaller velocity bins. Figure 2 shows an overlay of the CO J =
\low\ and J = \hig\ emissions in two different channel at velocities of
55.2 \kms\ (grey contours) and 50.1 \kms\ (black contours). It is clearly 
noted a separation of more than 20\arcsec\ between the 
emission at both velocities. Actually, the thinner layer located
toward the inner part of the nebula, at a velocity of $\sim 50$ \kms,
is adjacent to the more extended emission. The width of this layer
is of about $15-20\arcsec$, as we can see in the CO J\,=\,\hig\ map.

To better understand the velocity structure of the CO emission we have
analyzed the data along several position-velocity strips. The most 
illustrative one is shown in Fig.~3. This strip crosses both the
large and the thin molecular layers. It is clearly shown 
a sudden line broadening {\it just} at the position of the thin 
layer, around 50 \kms. This behaviour clearly demonstrates that the
feature noted in Fig.~2 is not merely due to a velocity gradient of a
single component, but to the appearing of a second component only
present at the interface between the molecular cloud and the optical 
nebula.

\subsection{Individual positions}

In order to have a complete picture of this striking feature, we performed
high-sensitivity, individual observations at the positions shown
as squares in Fig.~3a. These positions, labelled with letters from A to E,
are spaced by $\sim 20\arcsec$. Observed spectral lines at these 
locations are the CO J\,=\,\low, CO J = \hig, \co\ J\,=\,\low\, 
\co\ J\,=\,\hig, and C$^{18}$O J\,=\,\low\ 
lines. The \co\ J\,=\,\hig\ and C$^{18}$O J\,=\,\low\ lines were
detected for the first time in a W-R environment.

The resulting spectra are shown in Fig.~4. For all the observed lines, the 
peak velocity remains roughly constant from positions A to C, and the
linewidth increases. Both CO lines at position C are stronger than
at position B, while in both \co\ lines we note exactly the contrary. Toward
position D all the line intensities are roughly reduced to one half and a 
strong assymetry and line broadening appear. There is significant emission at
velocities as low as 47 \kms. The broadening shown in the position--velocity 
strip (Fig.~3b) is coincident with this position. C$^{18}$O is only
detected in positions B and C, corresponding to the highest emission at the
other isotopes.

Position E, although weak, shows a large CO J\,=\,\hig\ broadening to 
velocities as low as 42 \kms. This implies a total velocity interval 
emission of at least 14
\kms. It is worth mentioning that this very broad emission at \vlsr\ $<50$ 
\kms\ is not detected in the CO J = \low\ line. As mentioned above, we see 
in Fig.~3b a notable line broadening between positions C and D, where
the lower levels are seen at 47 and 57 \kms. Surprisingly, the high
sensitivity individual positions reveals a third velocity component at 
position E, which could not be 
detected using the \otf\ maps because of the sensitivity.

The main observational findings are summarized in Table 1. For each position,
this table shows the parameters of the five
observed transitions, obtained from a gaussian 
fit to a {\it single} component: antenna temperature of the 
maximum (T$_{\rm pk}$), LSR peak velocity (\vlsr) and full width at half 
maximum ($\Delta$V). The single component assumption is obviously not 
valid in some cases due to the strong assymetries of the line profiles, 
and hence is only provided for comparison.


\section{Physical parameters}

\subsection{Velocity components}

The observational features shown in Sect.~3 (in particular Figs.~3 and 4,
and Table 1) lead us to define three different velocity components.
The first component corresponds to the most extended molecular gas and is the
broad component, already studied in Paper I, at velocities from 52 to 
57 \kms. In the following, we shall refer to this component as the 
``low velocity component'' (\lvc). The higher spatial resolution of our maps 
allow us to disclose the presence of a second, less extended velocity 
component, seen as a thin layer ($\sim 20\arcsec$ width)
in Fig.~2, in the velocity range 48--52 \kms. We shall refer to this component
as the ``intermediate velocity component'' (\ivc). Finally, the high
sensitivity of the individual spectra revealed the third component, hereafter
referred to as the ``high velocity component'' (\hvc), as a very compact one,
clearly detected only in the CO J = \hig\ line toward position E (Fig.~4), at
velocities lower than 48 \kms. 

The extension in the sky corresponding to each component may be roughly 
inferred from Fig.~3(a): the \lvc\ is sketched in the greyscale, the \ivc\
is plotted in contours, and the \hvc\ is only detected in position E.
The \lvc\ corresponds to most of the gas detected in Paper I. It is the most 
extended component, and bounds the whole optical nebula by its southern and 
eastern borders. The \ivc\ is at the interface between \ngc\ and the 
most intense emission of \lvc. It seems to form a narrow layer between the \lvc\ 
and the optical nebula. Finally, the position where the \hvc\ was detected is 
projected onto the ionized gas and presumably has a compact morphology.

The stratified location of the three components with respect to the optical 
nebula \ngc\ and the quiescent molecular gas at 67 \kms, might be explained 
by succesive events associated to the evolution of a massive star. 
Therefore, different ages are expected for each component. The \lvc\ is the 
most extended, uniform and massive one, its velocity is the closest to the 
quiescent gas and hence is thought as the oldest one. The \ivc, however, is 
less extended than the \lvc\ and closer to the W-R star. These 
kinematical and morphological features might be explained by an event faster
and younger than that which formed the \lvc. So far, the \ivc\ may be the 
result of the encounter of two winds, in a manner suggested by 
Garc\'{\i}a-Segura et al.\ (1996a, 1996b). Finally, the \hvc\ may be 
recently shocked and would be the youngest component.

\subsection{Overall features}

Our \otf\ CO and \co\ maps trace the distribution of the molecular gas both 
into and around the southern optical nebula, including the interesting
molecular interface at its border. The different velocity components have 
probably been produced by violent energetic mechanisms usually linked to 
the massive star evolution. During the evolution of \hd, the stellar wind, 
eventual envelope ejections 
and the strong UV radiation fields could have produced 
shocked, ionized and photodissociated regions. The relative variations 
of different spectral lines should tell us about the physical conditions 
of the molecular gas.

We have analyzed the line ratios among the three observed spectral lines 
which covers the southern and eastern borders of \ngc.
In the following, we shall refer to the line ratio
of $^{12}$CO J = \low\ to \co\ J = \low\ as \risot, and
to the line ratio of $^{12}$CO J = \hig\ to $^{12}$CO J =
\low\ as \rexc. Both \risot\ and \rexc\ were computed for different
velocity ranges and for positions with line intensity 
greater than 3$\sigma$. \rexc\ was computed after convolving the CO
J = \hig\ map to the J = \low\ angular resolution. While \rexc\ is a good tracer
of the excitation temperature, \risot\ might trace the
combined effect of isotopic variations and changes in the CO opacity.

Figures 5a and 5b show the spatial distribution of \rexc\ for the velocity ranges
(52, 55) and (48, 51) \kms, roughly equivalent to the \lvc\ and the \ivc, 
respectively. For the \lvc\ (Fig.~5a), \rexc\ does not significantly changes in 
the brightest zone, and it is probably greater toward the south. A mean value 
for \rexc\ of
0.9 and 1.2 is found to the eastern and southern part of the map,
respectively. We have obtained similar results for the \ivc\ (Fig.~5b), 
where a mean value of 1 is found. However, the \co\ emission is really 
striking when compared to the
CO J = \low\ emission. Figure 6 sketches the distribution of \risot\ in the
velocity interval (52, 55) \kms\ (roughly the \lvc). It clearly changes
sistematically from southeast to northwest, and varies in almost one order of
magnitude. The sensitivity of our \otf\ maps were not enough to detect 
the \co\ emission in the velocity range corresponding to the \ivc, and 
\risot\ could not be computed.

\subsection{Excitation conditions}

From the overall information shown in Figs.~5 and 6, we see a rather
uniform excitation conditions over the whole region. So far, in most of the 
observed area, the excitation temperature does not varies significantly. 
However, large variations in the CO opacity might be inferred. If we assume 
a constant \co/CO isotopic abundance, the CO lines
will be optically thick in the eastern part of the cloud. This
feature explains the velocity difference between both isotopes, as noted 
in Paper I.

We have estimated the physical conditions for the three velocity 
components, by modelling the emission of the observed lines for positions
A to E, using the {\it large velocity gradient} method (LVG). The LVG 
method implements the Sobolev's formulation developed by Castor (1970)
and later extended by Jeffery (1995), among others.
We have computed the parameters assuming kinetic temperatures (\tk)
from 10 to 100\,K. For all components but the \hvc, a minimum \tk\ of 30\,K 
is needed to match the observed lines. This lower limit for \tk\ is in good
agreement with the results of Rizzo et al.\ (2001b), based on the metastable
NH$_3$ (1,1) and (2,2) lines detected at position B. Furthermore, the
results are not critically different for the temperature range 30--100\,K.
Both CO lines result to be optically thick in most cases, whereas 
both \co\ lines were optically thin in all cases. In consequence, we used 
the \co\ 2--1/1--0 line ratio (on a main-beam temperature scale) to estimate 
the H$_2$ density, n(H$_2$). For a given \tk, the
column densities of \co\ and C$^{18}$O were determined from the derived 
n(H$_2$) and the J = \low\ line intensity of each isotope. The CO column
density was estimated by assuming a constant isotopic ratio of 60 between
CO and \co\ (Wilson \& Matteucci 1992)

The Table 2 shows the results of applying the LVG method to each
component at each position, by assuming a \tk\ of 30\,K. By adopting this 
value for \tk\ we are in the most conservative case, because \tk\
may be higher in position E and probably in position D. Anyway, the
column densities do not vary in more than 10 percent (significantly smaller
than other sources of uncertainty) in this temperature range.
The first and second columns indicate the position and velocity component. 
The third column provides the obtained n(H$_2$). From fourth to sixth columns,
column density of CO, \co\ and C$^{18}$O are provided, respectively.

The results derived for \hvc\ in position E need to be explained in detail. 
This velocity component was only detected in the CO J = \hig\ line, and 
therefore has a {\it minimum} \rexc\ of 6. Such a high value could not 
be obtained having a \tk\ of 30\,K, even by considering that the emiting
region is unresolved by the 12\arcsec\ beam of the CO \hig\ line.
Therefore, it is neccesary larger values of \tk\ and 
a higher density to match the observed ratio. We estimate that \tk\ must be 
larger than 80\,K for this component, as well as the density should be at 
least $10^5$ cm$^{-3}$. The undetection of this component in the CO J = 
\low\  line strenghtens the idea of a small emiting volume, and a higher 
excitation temperature is also expected in this position (towards the inner 
part of the ionized nebula), as discussed in Paper I. Finally, it should be
noted that the CO column density for this component has been estimated from 
the CO J = \hig\ line.

\section{Origin and evolution of the CSM around WR\,7}
\subsection{Dynamics of the molecular gas around NGC\,2359}

The morphology and kinematics of each component, as well as their location 
with respect to \ngc\ and the quiescent molecular gas have been discussed 
in previous sections. Some of the physical conditions have been summarized 
in Table 2. In this section, we will put them together in order to draw a 
picture of the region and its possible evolution.

In Fig.~7, we plot the variation of the gas density
and \co\ column density as a function of the projected distance to the W-R 
star \hd, assumed at a distance of 5 kpc (Goudis et al.~1994). The 
plots correspond just to the \lvc\ and the \ivc\, and not to the \hvc, 
because it is only detected at a single position.
The stratification of these components are not only evident in the morphology 
and the kinematics. This progressive variation is also present in the maxium 
density (roughly 10$^3$ cm$^{-3}$ at \lvc, 10$^4$ cm$^{-3}$ at \ivc\ and at least 
10$^5$ cm$^{-3}$ at \hvc) and  in the total mass involved (regarding the column 
densities and the spatial extension of each component). If we take into account
the observed low values of \risot, it is also expected a trend in the opacity,
decreasing from \lvc\ to \hvc. The temperature is expected to 
increase toward the W-R star, and therefore in the sense \lvc--\ivc--\hvc.
For the \hvc, both the density and the temperature must be higher than in the 
other two components. Although a careful chemical study of the region is 
needed, we can associate each component to different evolutive episodes 
previous to the actual W-R status of \hd.

In Paper I, we have discovered a quiescent molecular cloud, that was 
interpreted as the original gas, which was later shocked by the expansion 
of a wind previous to W-R. 
The presence of molecular gas close to a W-R star is indeed puzzling, 
because it is located at a projected distance of only 
$\sim$ 8 pc, and has been exposed to the very strong radiation 
and stellar winds during all the evolutive phases of \hd. Therefore, the 
three velocity components reported here are the result of such interaction 
and represent the disturbed molecular gas linked to some of those phases.
In order to associate these features with the W-R phase or a previous one,
we have made crude estimates of some {\it global} dynamical parameters for 
each component, which are provided in Table 3. The whole nebula and its 
surroundings was completely 
mapped in Paper I, and we considered that the mass obtained there includes
both the \lvc\ and the \ivc. In this work, we roughly estimate that 80\%
of that mass are from the \lvc, while the other 20\% are from the \ivc.
The mass of the \hvc\ was estimated from the CO J = \hig\ spectra and 
assuming an angular extension of one beam. The computation of masses and 
projected distance were done by assuming a distance of 5 kpc (Goudis et al.~
1994). The expanding velocities were derived by assuming a rest velocity of 
67 \kms\ (Paper I). The dynamical time was computed according to Dyson 
(1989). Finally, Table 3 also includes estimates of kinetic energy and 
momentum of each component.

The mass is significantly different for each component, decreasing in the 
sense \lvc-\ivc-\hvc, while the other parameters are of the same order of
magnitude. All the components have dynamical ages of at most $10^5$ yr.
This time represents an upper limit of the age of the shock which produced 
the actual kinematics. It is one order of magnitude lower than the 
main-sequence {\hi} shell (2.3 
10$^6$, Paper I), and quite larger than the dynamical age of \ngc, 
1.3 10$^5$, determined by Treffers \& Chu (1982). However, it should be 
noted that this dynamical age for the optical nebula is indeed an upper 
limit, which is reflected by the 1.3--2.2 10$^4$ yr determined from Hipparcos 
proper motion for \hd\ between the center of the nebula and its actual position.
Due to the high symmetry of \ngc, we think that a few 10$^4$ yr represents 
a reasonable approach of the actual \ngc's age. In consequence, the observed 
CO features have been produced at an epoch between the end of the 
main sequence (MS) and the beginning of the W-R phase.

According to the current models, a W-R is the descendant of an O-star 
of at least 30 \msun. Depending on its initial mass, the O-star may evolve 
to W-R through different evolutive paths, including a red supergiant (RSG) or a 
luminous blue variable (LBV) phase. Other intermediate stages may
account, such as Of or different sub-classes of W-R (WNE, WNL, WC). Although
the MS phase is indeed a long-lived one, characterized by 
a strong stellar wind and an enormous rate of UV-radiation, we could not
simply relate the features observed in the molecular gas around \ngc\ to 
this phase, untill the other phases were also considered. Moreover, a MS
bubble around \hd\ was found in \hi\ (Paper I), and it has a lineal size of 
35-40 pc, closely similar to many other cases in the Galaxy, and in good
agreement with the theory.

It is indeed possible that the interaction of the molecular cloud with 
the MS wind could not be strong enough and the stellar wind could ran 
through the molecular cloud. This is a reasonable 
argument owing to the relatively low density of the MS wind. We now consider
the other evolutionary phases which may account for the observed features.
The more likely mechanism is that the stellar wind has accelerated and 
heated the quiescent molecular 
cloud, and hence it is more relevant the {\it instantaneous effect} than the 
accumulated one along the whole phase. Therefore, the phase responsible of
accelerating the gas should have a large momentum rate (\.P) and perhaps 
a large mechanical luminosity (\lmech). We also should pay attention to
the density of the stellar wind, because it is an indicator of how large
is, in comparison, the mass injected to the CSM at supersonic velocities.

\subsection{Models of stellar evolution}
Let us consider two different models of stellar evolution, which reproduce
two standard evolutive paths: (1) the 35 \msun\ model of Garc\'{\i}a-Segura 
et al.\ (1996b), with the path O$\rightarrow$RSG$\rightarrow$W-R; and (2) the 
60 \msun\ model of Langer et al.\ (1994), with the path 
O$\rightarrow$pre-LBV$\rightarrow$LBV$\rightarrow$W-R. We should note that 
some of these phases really includes several other, although no significant 
changes in the global stellar parameters are noted. So far, the W-R phase in 
both models can be divided in three sub-stages, such as a H-poor, 
a H-free and a WC stage.
In model (2), the pre-LBV phase includes both the Of and the H-rich WN phase.
The evolution of the CSM around both type of stars have
been modelled by Garc\'{\i}a-Segura et al.\ (1996a; 1996b).

The Table 4 provides mean representative values from each phase in each model,
such as the time spent, mass loss rate (\.M), \lmech, \.P, kinetic energy
E$_{\rm k}$ and total momentum (P) provided during the whole phase, and the 
wind density ($\rho_{\rm wind}$) at 5 pc from the star. The 35 \msun\
model is split in two cases, having a slow (15 \kms) and a fast (75 \kms) RSG 
wind.

\subsection{Acummulative effects}
In the three models, both the MS and the W-R stages provide the greatest
fraction of the total energy injected to the environs (10$^{50-51}$ erg). In 
the case of the MS 
stage, however, this is due to an accumulate effect, because the star spents
some Myr in MS, one or two orders of magnitude larger than the other phases.
In contrast, the W-R stage injects to the medium a similar amount of kinetic 
energy in some 10$^5$ yr, because its huge mass-loss rate. 
On the other hand, the {\it total} energy ouput associated to the RSG (fast 
wind) and LBV phases are two orders of magnitud (10$^{48}$ erg) lower than 
the MS or the W-R stages. The amount of kinetic energy associated to the RSG 
(slow wind) phase is even lower by about 4 orders of magnitude (10$^{46}$ erg).

A similar situation occurs when considering the total momentum provided in 
each phase, although the contrast is lower than in the energy.

\subsection{Instantaneous effects}
When we take into account the instantaneous effects (well described by 
\lmech, \.P\ and $\rho_{\rm wind}$), the situation changes dramatically. 
The intermediate 
stages RSG and LBV are characterized by an increase of \.M by {\it at least}
one order of magnitude with respect to their previous phases, and a sudden 
decrease of the terminal velocity. As a result, \lmech\ remains almost 
constant, while both \.P\ and $\rho_{\rm wind}$ increases significantly.

During these short-lived phases, the circumstellar gas is filled with a 
large amount of material ($\sim$ 20 \msun) from the stellar 
interior. This gas is relatively dense even at a few pc from the star and is
moving at low (although still supersonic) velocities. The RSG wind (slow wind)
is the densest one, but the momentum rate is practically the same as in the
MS case; so far, the effect onto the quiescent molecular cloud should not be
different from that of the MS stage. The RSG wind (fast wind) is somewhat higher, 
but not too different from MS wind; a factor of 2 or 3 in these simplified
estimates should not be seen as a significant one.

In contrast, the LBV wind is significantly different from the MS wind. The 
momentum rate is one order of magnitude greater than in any other evolutive stage,
and the wind density is three orders of magnitude larger than the previous 
phases (MS and pre-LBV) and two orders of magnitude larger than the W-R stage.

\subsection{Interpretation}

If the progenitor of the W-R star \hd\ has had a mass close to 35 \msun, the 
W-R wind is the unique source capable of disturbing the molecular cloud, 
because the RSG wind had not power enough to accelerate the original cloud.
So far, the W-R wind had encountered an already stopped dense RSG wind in 
contact with the molecular cloud. Consequently, the W-R wind has been suddenly 
decelerated and produced the observed shocked molecular gas.

If, on the other hand, the progenitor of \hd\ has had a higher mass, it is
very probable that the star had developed a LBV phase, which is characterized 
by a rather dense wind and a high \.P. The LBV wind might produce
what we are observing in CO. The values provided in Table 4 are {\it mean} values
for each phase as a whole. This simplification is good enough for all phases 
but LBV, because the parameters does not significantly change during the 
whole phase. However, the LBV phase consists of several episodes
of enlarged stellar winds (or mass ejections) produced by hydrodynamical 
instabilities in the stellar structure. It is thought that the mass-loss
rate during these episodes reaches values as high as several $10^{-3}$ \mlr,
increasing the mean values of \.M, \.P, \lmech\ and $\rho_{\rm wind}$ in 
Table 4 by a factor 10--20 during some $10^{3}$ yr. As in the 35 \msun\ case, 
the W-R wind
may have found the RSG wind and produced the shock observed around the W-R
nebula. The presence of {\it several} shockfronts (the three CO components and
the ionized layers into the optical nebula) strenghten this idea. If only 
the W-R stage had produced the observed features, it should have similar 
episodes of increased mass loss. 

The whole optical nebula would be associated to the W-R phase, and its
hot material (more or less uniform) would be located inside a multiple-layers 
wind with an earlier origin. In this case, the bounding of the nebula can be 
atomic or molecular, depending on the previous presence of a molecular cloud 
close to the star. The morphology of the molecular gas in \ngc, as 
well as in NGC\,6888 and the nebula around WR\,134 (Rizzo et al.\ 2003a) are 
strongly suggestive in this sense.

In any case, it is clear the need of enlarge the sample, search for possible 
chemical differences which may confirm our hypothesis, observe and characterize
the stellar winds of the well-known LBVs (P Cyg, $\eta$ Car, HR Car) and construct
hydrodynamical models which take into account the different evolutive phases 
{\it and the presence of a close molecular cloud}.


\section{Future perspectives}

The history of a massive star is written by the impacts onto the surrounding
gas, particularly in the molecular component. Therefore, we can increase our
knowledge about massive stars evolution by observing the consequences in their 
environments. There are just a few observed cases of interaction between W-R
stars and the surrounding molecular gas (Marston et al.\ 1999;  
Paper I; Cappa et al.\ 2001; Rizzo et al.\ 2003a). However, we think 
that \ngc\ becames the best studied case, where a time evolution
may be followed and where high-density gas has been detected. New
observational efforts are neccesary in order to increase the sample and 
confirm some of the results obtained here and predicted by the theory.

The increasing detection of molecular gas around W-R nebulae opens questions 
which are not completely fulfilled by current models. In
the future, new hydrodynamical simulations should address this aspect of the
interplay between evolved massive stars and their environs. It is expected to 
know where and when the molecules can be formed or can survive around these 
hot massive stars.

On the other hand, the detection and chemical analysis of complex molecules
in these ambients, at mm and sub-mm lines, would tell us about the chemical 
composition and the excitation conditions of the gas. In this sense, just 
a few detections have been reported (Marston 2001; Rizzo et al.~2001b, 
2003b). This is a key subject, because we can learn not merely about the 
interplay of the massive star evolution with its surroundings, but also the 
ambient where a supernova takes place. So far, this field is also connected 
with the chemical evolution of the Galaxy and its energetic balance.


\section{Conclusions}

As shown in Paper I, the southeastern border of \ngc\ contains the most clear
sign of interaction of a W-R nebula with the molecular environment. The 
original molecular cloud was partially accelerated by at least 14 \kms\ (from
67 to 53 \kms) during one of the early evolutive phases of \hd. In this work, 
we have used the \otf\ tecnique to obtain detailed maps of the CO and \co\ 
emission of this disturbed molecular gas. The angular resolution and 
sensitivity achieved allowed us to distinguish three velocity components, 
clearly stratified with respect to the nebula and the quiescent gas.

The first component (called the \lvc\ throughout the paper) is the most 
extended and the most massive one. The second component, the \ivc, appears
as a thin layer (less than 1 pc thickness) located at the interface between the
\lvc\ and the optical nebula. Specially striking is the third component, the 
\hvc, which appears as a
very compact component, only detected in the CO J = \hig\ line towards the 
inner part of the optical nebula. The \lvc\ has LSR velocities between 52 and
57 \kms\ (the nearest to the quiescent gas), the \ivc\ appears at 48--52 
\kms, and the \hvc\ emits at velocities between 42 and 48 \kms. So far, the
stratification of these components are both morphological and 
kinematical features.

By observing single positions with higher sensitivity along a strip --more 
or less radially with respect to WR\,7--, some physical parameters 
(volume density, column density and opacity) have been inferred. The density 
increases in the sense \lvc--\ivc--\hvc\ from 10$^3$ cm$^{-3}$ to at least 
10$^5$ cm$^{-3}$, while the column density, total mass and 
opacity should decrease in the same sense. We have proposed a scenario 
where the three components are produced during different evolutive episodes 
of the predecessors of \hd. Each component would be the result of an encounter 
(likely shocks) of the expanding stellar winds (or mass ejections) with 
the material ejected during the previous 
phase or even with the ambient gas. This idea qualitatively fits some 
theoretical results (Garc\'{\i}a-Segura et al.\ 1996a, 1996b), although some
specific aspects should be analyzed in detail. Particularly, the possible
presence of molecular gas at a few pc from the massive star should be 
considered in future theoretical works.

\acknowledgements
This work was partially supported by the Ministerio de Ciencia y
Tecnolog\'{\i}a of Spain (MCyT), by the grant AYA 2003-06473.
JM-P acknowledges the financial support from the MCyT grants ESP 
2002-01627 and AYA 2002-10113E. JRR acknowledges
Dr. M.\ A.\ G\'omez-Flechoso for very profitable talks about
the theoretical models discussed in the paper. The authors are
specially grateful with the referee comments, which helped to improve 
the paper.



\clearpage

\begin{table*}
\caption[]{Gaussian fits to the observed CO isotopic spectral lines}
\begin{tabular}{
c@{\qquad}
c@{\quad}c@{\quad}cc@{\qquad}
c@{\quad}c@{\quad}cc@{\qquad}
c@{\quad}c@{\quad}cc@{\qquad}
c@{\quad}c@{\quad}cc@{\qquad}
c@{\quad}c@{\quad}c
}
\hline
\hline
\noalign{\smallskip}

Pos. &
\multicolumn {3}{c}{CO J = 1$\rightarrow$0} & &
\multicolumn {3}{c}{CO J = 2$\rightarrow$1} & &
\multicolumn {3}{c}{$^{13}$CO J = 1$\rightarrow$0} & &
\multicolumn {3}{c}{$^{13}$CO J = 2$\rightarrow$1} & &
\multicolumn {3}{c}{C$^{18}$O J = 1$\rightarrow$0} \\

\noalign{\smallskip}

&
T$_{\rm pk}$ & V$_{\rm LSR}$ & $\Delta$V & &
T$_{\rm pk}$ & V$_{\rm LSR}$ & $\Delta$V & &
T$_{\rm pk}$ & V$_{\rm LSR}$ & $\Delta$V & &
T$_{\rm pk}$ & V$_{\rm LSR}$ & $\Delta$V & &
T$_{\rm pk}$ & V$_{\rm LSR}$ & $\Delta$V \\

\noalign{\smallskip}

&
K & \multicolumn{2}{c}{km s$^{-1}$} &&
K & \multicolumn{2}{c}{km s$^{-1}$} &&
K & \multicolumn{2}{c}{km s$^{-1}$} &&
K & \multicolumn{2}{c}{km s$^{-1}$} &&
K & \multicolumn{2}{c}{km s$^{-1}$} \\

\noalign{\smallskip}
\hline
\noalign{\smallskip}

A&
2.6 & 53.4 & 2.1 && 2.6 & 53.3 & 2.3 && 0.5 & 53.8 & 1.9 && 0.4 & 53.2 & 2.2 &&
\multicolumn{3}{c}{$<$ 0.04 K km s$^{-1}$ $^{(a)}$} \\

B&
6.3 & 53.7 & 2.8 && 4.4 & 53.6 & 3.1 && 1.5 & 54.0 & 2.1 && 1.8 & 53.7 & 1.9 &&
0.08 & 53.6 & 2.6 \\

C&
7.4 & 53.3 & 4.2 && 5.4 & 53.0 & 4.7 && 1.0 & 53.9 & 2.7 && 0.8 & 53.2 & 4.0 && 0.08 & 53.7 & 2.8 \\

D&
3.6 & 52.3 & 5.3 && 2.6 & 51.8 & 5.5 && 0.3 & 52.5 & 5.8 && 0.2 & 52.0 & 5.2 &&
\multicolumn{3}{c}{$<$ 0.09 K km s$^{-1}$ $^{(a)}$} \\

E&
0.9 & 53.2 & 4.8 && 0.6 & 53.0 & 14.0 && 0.1 & 53.7 & 4.3 && 0.1 &
53.4 & 1.5 && \multicolumn{3}{c}{$<$ 0.05 K km s$^{-1}$ $^{(a)}$} \\

\noalign{\smallskip}
\hline
\noalign{\smallskip}
\end{tabular}
\\

$^{(a)}$ Upper limit of the integrated line intensity.\\


\end{table*}

\clearpage

\begin{table}
\caption[]{LVG results for the CO isotopes in \ngc}
\begin{tabular}{
l@{\quad}l@{\quad}
c@{\quad}c@{\quad}c@{\qquad}
c@{\quad}c@{\quad}c
}
\hline
\hline
\noalign{\smallskip}

Pos. & Comp. & n(H$_2$) & N(CO) & N($^{13}$CO) & N(C$^{18}$O) & 
$^{13}$CO/C$^{18}$O\\

\noalign{\smallskip}

& & 10$^3$ cm$^{-3}$ & 10$^{16}$ cm$^{-2}$ & 10$^{14}$ cm$^{-2}$ & 
10$^{14}$ cm$^{-2}$ & \\

\noalign{\smallskip}
\hline
\noalign{\smallskip}

A & \lvc\ & 3.0	 & 6.1  & 10.2 & $<0.8$ & $>12.9$ & \\
B & \lvc\ & 4.9	 & 24.4 & 40.7 & 2.2 & 18.2 \\
C & \lvc\ & 3.0	 & 17.3 & 28.8 & 1.9 & 14.8 \\
C & \ivc\ & 11.5 & 3.8  & 6.3  & $<1.5$ & $>4.3$ \\
D & \lvc\ & 3.9	 & 3.5  & 5.9  & $<1.3$ & $>4.4$ \\
D & \ivc\ & 3.0	 & 5.3  & 8.9  & $<1.4$ & $>6.5$ \\
E & \lvc\ & 1.4	 & 1.7  & 2.8  & $<0.8$ & $>3.5$ \\
E & \ivc\ & 1.7	 & 0.2  & $<0.7$  & $<1.0$ & \nodata \\
E & \hvc\ & $\ga 100$ & 0.2 & $<2.0$ & $<1.6$ & \nodata \\

\noalign{\smallskip}
\hline
\noalign{\smallskip}
\end{tabular}
\\


\end{table}

\clearpage

\begin{table}
\caption[]{Global parameters of the molecular layers in \ngc}
\begin{tabular}{lcccccc}
\hline
\hline
\noalign{\smallskip}

Comp. & Mass & V$_{\rm exp}$ & R$_{\rm proj}$ & T$_{\rm dyn}$ & 
E$_{\rm k}$ & P \\

\noalign{\smallskip}

& \msun & \kms & pc & yr & erg & \msun\ \kms \\

\noalign{\smallskip}
\hline
\noalign{\smallskip}

\lvc & 130 & 12 & 6.5 & $3.0 10^5$ & $2.0 10^{47}$ & $1.7 10^3$ \\ 
\ivc &  30 & 17 & 5.8 & $1.9 10^5$ & $1.2 10^{47}$ & $6.8 10^2$ \\
\hvc &   8 & 23 & 5.1 & $1.2 10^5$ & $4.8 10^{46}$ & $2.1 10^2$ \\

\noalign{\smallskip}
\hline
\noalign{\smallskip}
\end{tabular}
\\


\end{table}


\clearpage

\begin{table*}
\caption[]{Evolutive phases of massive stars}
\begin{tabular}{lccccccc}
\hline
\hline
\noalign{\smallskip}

Phase & Time & \.M & \lmech\ & \.P & E$_{\rm k}$ & P & $\rho_{\rm wind}$\\

\noalign{\smallskip}
& yr & \mlr & erg & \msun\ \kms\ yr$^{-1}$ & erg & \msun\ \kms & cm$^{-3}$\\
\noalign{\smallskip}
\hline
\\
\multicolumn{6}{l}{35 \msun, slow wind. Model from Garc\'{\i}a-Segura et al.\ 1996b}\\
O-star & $4.5 10^{6}$ & $5.6 10^{-7}$ & $1.1 10^{36}$ & $1.4 10^{-3}$ & 
$1.6 10^{50}$ & $6.3 10^{3}$ & $2.4 10^{-5}$ \\
RSG    & $2.3 10^{5}$ & $8.1 10^{-5}$ & $5.8 10^{33}$ & $1.2 10^{-3}$ & 
$4.2 10^{46}$ & $2.8 10^{2}$ & $6.8 10^{-1}$ \\
W-R    & $1.9 10^{5}$ & $2.4 10^{-5}$ & $3.4 10^{37}$ & $5.0 10^{-2}$ & 
$2.0 10^{50}$ & $9.6 10^{3}$ & $1.2 10^{-3}$ \\
\\

\multicolumn{6}{l}{35 \msun, fast wind. Model from Garc\'{\i}a-Segura et al.\ 1996b}\\
O-star & $4.5 10^{6}$ & $5.6 10^{-7}$ & $1.1 10^{36}$ & $1.4 10^{-3}$ & 
$1.6 10^{50}$ & $6.3 10^{3}$ & $2.4 10^{-5}$ \\
RSG    & $2.3 10^{5}$ & $8.1 10^{-5}$ & $1.4 10^{35}$ & $6.1 10^{-3}$ & 
$1.1 10^{48}$ & $1.4 10^{3}$ & $1.4 10^{-1}$ \\
W-R    & $1.9 10^{5}$ & $2.4 10^{-5}$ & $3.4 10^{37}$ & $5.0 10^{-2}$ & 
$2.0 10^{50}$ & $9.6 10^{3}$ & $1.2 10^{-3}$ \\
\\

\multicolumn{6}{l}{60 \msun. Model from Langer et al.\ 1994}\\
O-star & $1.7 10^{6}$ & $4.7 10^{-6}$ & $1.0 10^{37}$ & $1.2 10^{-2}$ & 
$5.4 10^{50}$ & $2.1 10^{4}$ & $2.5 10^{-4}$ \\
pre-LBV& $1.5 10^{6}$ & $1.5 10^{-5}$ & $2.7 10^{37}$ & $3.6 10^{-2}$ & 
$1.4 10^{51}$ & $5.8 10^{4}$ & $9.2 10^{-4}$ \\
LBV    & $1.2 10^{4}$ & $6.7 10^{-4}$ & $2.6 10^{37}$ & $2.4 10^{-1}$ & 
$9.9 10^{48}$ & $2.8 10^{3}$ & $2.4 10^{-1}$ \\
W-R    & $6.5 10^{5}$ & $2.4 10^{-5}$ & $3.0 10^{37}$ & $4.8 10^{-2}$ & 
$6.2 10^{50}$ & $3.1 10^{4}$ & $2.0 10^{-3}$ \\
\\

\noalign{\smallskip}
\hline
\noalign{\smallskip}
\end{tabular}
\\

\end{table*}

\clearpage
\begin{figure*}
{\large Figure captions}\\
\caption{
CO and \co\ integrated emission from 50 to 56 \kms\ toward \ngc.
{\bf a)} J-band picture of the optical nebula, overlaid
with the CO J = \hig\ emission from Rizzo et al.\ (2001a). The
rectangle indicates the region mapped in this paper with higher
angular resolution. {\bf b)}, {\bf c)} and {\bf d)} Molecular transitions
indicated at the top left of each map, with the beams plotted in
the bottom left corner. Angular offsets in Figs.\ {\bf b)}, {\bf c)} and {\bf
d)} are referred to (RA, Dec.)$_{2000}$ =
($07^{\rm h}18^{\rm m}35\fs7,-13\degr16\arcmin37\farcs0$).
First level and spacing in these figures are 4.4, 0.8 and 4
K\,\kms, respectively.
}
\end{figure*}

\begin{figure*}
\caption{
Comparison of the CO emission in two velocity channels. {\bf a)} CO 
J = \low\ brightness temperature at 55.2 \kms\ (grey contours),
overlaid with the emission at 50.1 \kms\ (black contours). Lower contour
and step are 0.3 K (grey) and 0.l6 K (black). {\bf b)} The same as {\bf a)}
for the CO J = \hig\ brightness temperature. Lower contour and step are
0.15 K (grey) and 0.8 K (black). We note in both maps a clear shift in the
emission, separated by $\sim 20 \arcsec$.
}
\end{figure*}

\begin{figure*}
\caption{
A position-velocity strip in the CO J = \hig\ emission. {\bf a)} 
CO J = \hig\ emission map integrated between 52 and 55 \kms\ (on 
greyscale, from 2 to 22 K\,\kms) and between 48 and 51 \kms\ 
(contours, 1 to 5 times 0.2 K\,\kms). The straight line indicates
the location of the strip. The filled squares along 
the line show the 
individual positions observed with higher sensitivity (see Sect.~3.2 
and Fig.~4). The three kinematical components described in Sect.~4 
are distinguished, since the greyscale represents the {\it low velocity
component}, the contours sketches the {\it intermediate velocity 
component} and the {\it high velocity component} is detected just in
position E. {\bf b)} 
Position-velocity strip in the velocity range of interest. Angular
distance is measured from southeast to northwest. The bottom small triangles
indicate the approximate location of the individual positions marked by 
squares in {\bf a)}. Contours are 1 to 10 times 0.6 K. In general, we do not 
see a velocity variation greater than 1 \kms\ along the slice. However, a 
striking line broadening toward
the inner part of the nebula appears, in close coincidence with the thin
layer at 48--50 \kms\ depicted in Fig.~2.
}
\end{figure*}

\begin{figure*}
\caption{
High-sensitivity, individual spectra of CO isotopes along the strip shown
in Fig.~3. Every column corresponds to the spectral line written at the
top, while every row corresponds to the selected positions. The assymetries
in positions C, D and E indicates the presence of different velocity
components. It is striking the very broad emission of the CO J = \hig\ line
in position E. \co\ J = \hig\ and C$^{18}$O J = \low\ are detected for the
first time in a W-R nebula. Temperature scale is T$_{\rm A}^*$ (K). Since 
different scales are used, auxiliary lines at 1 K (for both CO lines) and
0.2 K (for the $^{13}$CO lines) were also plotted.
}
\end{figure*}

\begin{figure*}
\caption{
CO J = \hig\ to J = \low\ line ratio (\rexc) in the southern part of \ngc. The
velocity ranges are indicated at the left of each figure. {\bf a)} ``low 
velocity component''. {\bf b)} ``intermediate velocity component''. In
general, \rexc\ does not show no large changes in both velocity intervals, 
except toward the inner nebula, in panel {\bf a)}, where \rexc\ is at
least a factor of six larger than the mean value.
}
\end{figure*}

\begin{figure*}
\caption{
CO to \co\ J = \low\ line ratio (\risot) in the southern part of \ngc. The velocity
range of integration is (52, 55) \kms. A systemic change (almost by one order
of magnitude) from east to west is noted all over the map. This
might indicate a different distribution of CO and \co, variation of abundance
or opacity effects in the area.
}
\end{figure*}

\begin{figure*}
\caption{
Variation of some physical parameters for the low (\lvc; filled squares) 
and intermediate (\ivc; filled triangles) velocity components, as a 
function of the distance to \hd. {\bf a)} Distribution 
of the density (in logarithmic units). {\bf b)} \co\ 
column density. Abcissae are projected distance to the W-R star by assuming 
a distance of 5 kpc. Open symbols represent upper limits.
}
\end{figure*}

\end{document}